\documentclass[aps,prc,superscriptaddress]{revtex4}
\usepackage{amsmath}
\usepackage{graphicx}
\usepackage{hyperref}

\begin{document}
\title{The shear viscosity of parton matter under anisotropic scatterings}
\author{Noah M. MacKay, Zi-Wei Lin}
\affiliation{Department of Physics, East Carolina University,
  Greenville, North Carolina 27858, USA}
\date{\today}

\begin{abstract}
The shear viscosity $\eta$ of a quark-gluon plasma in equilibrium can
be calculated analytically using multiple methods or numerically using
the Green-Kubo relation. It has been realized, which we confirm here,
that the Chapman-Enskog method agrees well with the Green-Kubo result
for both isotropic and anisotropic two-body scatterings.  We then apply the
Chapman-Enskog method to study the shear viscosity of the parton
matter from a multi-phase transport model. 
In particular, we study the parton matter in the center 
cell of central and midcentral Au+Au collisions at $200A$ GeV and Pb+Pb
collisions at $2760A$ GeV, which is assumed to be a plasma in thermal
equilibrium but partial chemical equilibrium. 
As a result of using a constant Debye mass
or cross section $\sigma$ for parton scatterings, the $\eta/s$ ratio
increases with time (as the effective temperature  decreases),
contrary to the trend preferred by Bayesian analysis of the
experimental data or pQCD results that use temperature-dependent
Debye masses. At $\sigma=3$ mb that enables the transport model to
approximately reproduce the elliptic flow data of the bulk matter, the
average $\eta/s$ of the parton matter in partial equilibrium is
found to be very small, between one to two times $1/(4\pi)$. 
\end{abstract}
\maketitle 

\section{Introduction}

In ultra-relativistic heavy ion collisions at the Relativistic Heavy
Ion Collider (RHIC) and the Large Hadron Collider (LHC), a quark-gluon
plasma (QGP) has been created~\cite{STAR:2005gfr,PHENIX:2004vcz}. The
comparisons~\cite{Romatschke:2007mq,Song:2008si,ALICE:2010suc} between
the experimental measurements of the anisotropic flows and theoretical
models such as hydrodynamics suggest that the QGP behaves like a
near-perfect fluid with a very  small $\eta/s$ (shear viscosity to  
entropy density) ratio, not too far from the lower bound $1/(4\pi)$
from the conformal field theory~\cite{Kovtun:2004de}. 

Similar to hydrodynamics models, transport models can also describe
the large observed elliptic flow in high energy heavy ion collisions
once parton interactions are
included~\cite{Lin:2001zk,Xu:2007jv,Xu:2008av,Ferini:2008he}. 
In these transport models, the interactions among partons are
typically represented by the interaction cross section(s), including
the magnitude and angular distribution, which then determine the
plasma properties including the shear viscosity $\eta$ and the heavy
quark spatial diffusion coefficient $D_s$.  Unlike hydrodynamics
models, where the $\eta/s$ ratio (including its possible temperature
dependence) is an input parameter,  transport model calculations can
only be related to the QGP shear viscosity or $\eta/s$  after applying
the relation between the parton cross section(s) and shear
viscosity~\cite{Xu:2007jv,Xu:2007ns,Ferini:2008he,Xu:2011fi}.

The shear viscosity of a parton matter at a given temperature can be
calculated analytically or numerically. Analytical methods include
the Israel-Stewart (IS),  Navier-Stokes (IS), relaxation time
approximation (RTA), and Chapman-Enskog (CE)~\cite{chapman,deGroot}
methods. They often give different results, especially for anisotropic
scatterings. The shear viscosity can also be numerically calculated
with the Green-Kubo
relation~\cite{Muronga:2003tb,Demir:2008tr,Fuini:2010xz}.  An earlier
study~\cite{Plumari:2012ep} has found that for two-body anisotropic
scatterings the Chapman-Enskog method agrees well with  the Green-Kubo
result while the relaxation time approximation (even a modified
version RTA$^*$) does not.
 
In this study, we first examine the four analytical methods (IS, NS,
RTA$^*$, CE) in comparison with the Green-Kubo numerical results on
the shear viscosity and $\eta/s$ ratio for massless partons in
equilibrium under isotropic or  anisotropic two-body scatterings. 
We then apply the CE method to the parton matter in the string melting
version of a multi-phase transport (AMPT) model~\cite{Lin:2014tya},
which we assume as a QGP in partial equilibrium. In particular, we
study the time evolution of $\eta$ and $\eta/s$ of the center cell of
the parton matter in high energy Au+Au and Pb+Pb collisions. After
some discussions, we then summarize our findings. 

\section{Methods}

In addition to the isotropic scattering cross section, we also
consider the forward-angle scattering cross section, which is 
based on the perturbative QCD (pQCD) $gg\rightarrow gg$ cross section and
implemented in the ZPC~\cite{Zhang:1997ej} and AMPT~\cite{Lin:2004en}
models. The differential cross section for forward scattering is given
by~\cite{Lin:2004en}
\begin{equation}
\frac{d\sigma}{d\hat{t}}=\frac{9\pi\alpha_s^2}{2}\left(1+\frac{\mu^2}{\hat{s}}\right)\frac{1}{(\hat{t}-\mu^2)^2}, 
\label{dsdt}
\end{equation}
where $\hat s$ and $\hat t$ are the standard Mandelstam variables,
$\alpha_s$ is the strong-coupling constant, and $\mu$ is the Debye
screening mass. The factor $(1+\mu^2/{\hat s})$ is used to make the
total cross section energy independent (i.e., independent of $\hat
s$): 
\begin{equation}
\sigma= \frac{9\pi\alpha_s^2}{2\mu^2}.
\end{equation}

The transport cross section, defined as~\cite{Molnar:2001ux} 
\begin{equation} 
\sigma_{\mathrm{tr}}=\int d\sigma \sin^2\theta_{\mathrm{cm}},
\label{trans}
\end{equation}
where $\theta_{\mathrm{cm}}$ is the scattering angle in the two-parton
center of mass frame, often appears in the viscosity expressions because
the shear viscosity depends on the effectiveness of momentum transfer
of the parton scatterings. For the above forward cross section, one 
gets~\cite{Molnar:2001ux,Plumari:2012ep} 
\begin{equation}
\sigma_{\mathrm{tr}}=4 a (1+a) \left[ (1+2a) \ln \left (1+\frac{1}{a}
  \right) -2 \right] \sigma  \equiv h(a) \; \sigma, 
\end{equation}
where $a=\mu^2/\hat{s}$. As $a \rightarrow 0$, the cross
section becomes more forward-peaked and $h(a) \rightarrow 0$.  On the
other hand, as $a \rightarrow \infty$, the differential cross section
of Eq.\eqref{dsdt}  becomes isotropic and $h(a) \rightarrow 2/3$. 
Also note that the $h(a)$ function increases monotonously with $a$. 
Therefore, $h(a)$ is directly related to the anisotropy of the
scattering cross section. Since the transport cross section depends on
$\hat s$ even if the total cross section does not, we take its
thermal average~\cite{Kolb:1983sj}  when considering a parton matter
at a given temperature $T$:
\begin{equation}
\langle \sigma_{\mathrm{tr}}\rangle=\frac{\sigma}{32}\int_0^\infty du~
h\! \left (\frac{w^2}{u^2} \right) \left[ u^4K_1(u)+2u^3K_2(u)\right]
\equiv \sigma g_0(w).
\label{sigtr}
\end{equation}
In the above, $w=\mu/T, u=\sqrt{\hat s}/T$ is the integration variable,
$K_n$ is the modified Bessel function of the second kind,  
and Boltzmann statistics is used for the thermal distribution. 
The $g_0(w)$ function defined above is just the thermal average of
$h(a)$, it thus approaches 0 as $w \to 0$ but approaches 2/3 as $w \to
\infty$. 

Next we examine four analytical methods and the numerical Green-Kubo
method for calculating the shear viscosity of a parton matter under
isotropic or forward scatterings. 
Note that we only consider a massless QGP. Therefore, the  entropy
density is given by  $s=4 g_{\rm B}T^3/\pi^2$, where $g_{\rm B}=4(4 +
3N_f )$ is the total degeneracy factor of the QGP with Boltzmann
statistics, and $N_f$ is the number of relevant quark flavors.

\subsection{Israel-Stewart and Navier-Stokes Methods}

The Israel-Stewart~\cite{is1,is2} and Navier-Stokes~\cite{deGroot}
expressions for shear viscosity can be written respectively as    
\begin{equation} 
\eta^{\,\mathrm{IS}}=\frac{6~T}{5~\sigma},
~\eta^{\,\mathrm{NS}}\simeq 1.2654 \frac{T}{\sigma} 
\label{isoeta}
\end{equation}
for isotropic scatterings, where $\sigma_{\mathrm{tr}}=2\sigma/3$. 
They can be generalized to anisotropic scatterings as~\cite{Huovinen:2008te}
\begin{equation}
\eta^{\,\mathrm{IS}}=\frac{4~T}{5\langle\sigma_{\mathrm{tr}}\rangle},
~\eta^{\,\mathrm{NS}}\simeq 0.8436\frac{T}{\langle \sigma_{\mathrm{tr}}\rangle}.
\label{anisoeta}
\end{equation}

\subsection{The Modified Relaxation Time Approximation}

The relaxation time approximation is widely used in kinetic theory, 
where one approximates the collision integral in the Boltzmann equation as 
$\mathcal{C}[f]\propto -(f-f_{eq})/\tau$. Here $f$ is the particle
distribution function with $f_{eq}$ being the one in equilibrium, and
$\tau$ is the relaxation time.   With Boltzmann statistics, the RTA
expression for the shear viscosity is given by~\cite{aw} 
\begin{equation}
\eta^{\mathrm{RTA}}=\frac{4~T}{5~\sigma}.
\label{etaRTA}
\end{equation}

In the modified RTA method (denoted as RTA$^*$ in this
study)~\cite{Plumari:2012ep}, Eq.\eqref{etaRTA} is changed to the
following: 
\begin{equation}
\eta^{\mathrm{RTA^*}}=\frac{4~T}{5\langle \sigma_{\mathrm{tr}}
  v_{\mathrm{rel}}\rangle}.
\label{etaMRTA}
\end{equation}
In the above, $v_{\mathrm{rel}}=\sqrt{ {\hat s} ({\hat s}-4m^2)}
/(2E_1E_2)$ is the relative velocity between the two colliding
partons, where $E_1$ and $E_2$ represent respectively the energy of
the  two partons (each with mass $m$); for massless partons one has
$\langle v_{\mathrm{rel}}\rangle=1$. 
Note that for isotropic scatterings this modification 
gives $\eta^{\mathrm{RTA^*}}=6T/(5\sigma)$, which fails to reproduce
the original RTA result of Eq.\eqref{etaRTA}; however, it agrees with
the Israel-Stewart  (and Chapman-Enskog) expression. 
For anisotropic scatterings, it differs from the  Israel-Stewart
expression by a $v_{\mathrm{rel}}$ term, with the thermal average 
given by~\cite{Plumari:2012ep,Koch:1986ud} 
\begin{equation}
\langle \sigma_{\mathrm{tr}}v_{\mathrm{rel}} \rangle
=\frac{8z}{K_2^2(z)}\int_1^\infty dy~y^2(y^2-1)K_1(2zy)
\int d\sigma \sin^2\theta_{\mathrm{cm}}, 
\label{sigtrv1}
\end{equation}
where $z=m/T$ and $y=\sqrt{\hat{s}}/(2m)$. 
For massless partons and the AMPT differential cross section of
Eq.\eqref{dsdt}, we obtain 
\begin{equation}
\langle \sigma_{\mathrm{tr}}v_{\mathrm{rel}}
\rangle=\frac{\sigma}{16}\int_0^\infty du~
h\!\left(\frac{w^2}{u^2}\right)u^4K_1(u) \equiv \sigma g_1(w),  
\label{sigtrv2}
\end{equation}
where $g_1(w)$ is effectively another thermal average of $h(a)$.

\subsection{The Chapman-Enskog Method}

The Chapman-Enskog method solves the Boltzmann equation by applying a
series expansion on the distribution function~\cite{chapman}.  The
first-order result for the general case of massive particles under an 
anisotropic cross section is given by~\cite{Wiranata:2012br} 
\begin{equation}
\eta^{\mathrm{CE}}=\frac{T}{10}\frac{\gamma_0^2}{c_{00}},
\label{cemass1}
\end{equation}
where $\gamma_0=-10K_3(z)/K_2(z)$, and
\begin{equation}
\frac{c_{00}}{\gamma_0^2}=\frac{4z^3}{25K_3^2(z)}\int_1^\infty
dy~(y^2-1)^3\left[\left(y^2+\frac{1}{3z^2}
  \right)K_3(2zy)-\frac{y}{z}K_2(2zy)\right] \int d\sigma
\sin^2\theta_{\mathrm{cm}}. 
\label{cemass2}
\end{equation}
For the special case of massless partons and the AMPT differential
cross section of Eq.\eqref{dsdt}, we then obtain~\cite{Plumari:2014fda} 
\begin{equation} 
\frac{c_{00}}{\gamma_0^2}=\frac{\sigma}{51200}\int_{0}^\infty du~
h\!\left(\frac{w^2}{u^2} \right) u^6 \left[\left(\frac{u^2}{4} +\frac{1}{3}
  \right)K_3(u)-\frac{u}{2}K_2(u) \right]
\equiv \frac{\sigma}{8} g_2(w). 
\label{ce1}
\end{equation}
For isotropic scatterings, $g_2(w)=2/3$ and
$\eta^{\mathrm{CE}}=\eta^{\mathrm{IS}}$. 
For anisotropic scattering in general, we have 
\begin{equation} 
\eta^{\mathrm{CE}}=\frac{4~T}{5~\sigma g_2(w)}.
\label{ce2}
\end{equation}
Comparing to $\eta^{\mathrm{IS}}$ in Eq.\eqref{anisoeta}, we see that 
$\sigma g_2(w)$ in the CE method serves the role of 
$\langle\sigma_{\mathrm{tr}}\rangle$ in the IS method, 
and $g_2(w)$ is another thermal average of $h(a)$ similar to $g_0(w)$
in Eq.\eqref{sigtr} and $g_1(w)$ in Eq.\eqref{sigtrv2}.

The $\eta$ expressions in Eqs.\eqref{cemass1}-\eqref{cemass2} for the
general case of massive partons were given
earlier~\cite{Wiranata:2012br,Plumari:2012ep}, while
Eqs.\eqref{ce1}-\eqref{ce2} for the special massless case were shown
later~\cite{Plumari:2014fda}.  Note that there is a typo in  
Eq.(35) and Eq.(38) of Ref.~\cite{Plumari:2012ep}, which give the
$\eta$ result for the modified RTA and Chapman-Enskog methods
respectively, where $h(2zy\bar a)$ in the two equations should be
$h(a)=h(1/(2zy\bar a)^2)$ since $2zy\bar a=\sqrt{\hat s}/\mu \neq a$ 
with $\bar a \equiv T/\mu$~\cite{Plumari:2012ep}.

\subsection{The Green-Kubo Relation}

The Green-Kubo relation~\cite{Green,Kubo} can be used to numerically
calculate the shear viscosity at or near
equilibrium~\cite{Muronga:2003tb,Demir:2008tr,Fuini:2010xz}. 
In an earlier work~\cite{Zhao:2020yvf}, we have calculated the shear
viscosity of a massless gluon gas in equilibrium in a box under
isotropic or anisotropic two-body scatterings according to the
following form of the Green-Kubo relation~\cite{Muronga:2003tb}: 
\begin{equation}
\eta =\frac{V}{T} \int_{0}^{\infty}dt \,
\langle \bar{\pi }^{xy}(t+t^\prime)\,\bar{\pi }^{xy}(t^\prime) \rangle.
\label{gk}
\end{equation}
In the above, $V$ is the volume of the gluon gas, and the bracket 
represents the time ($t^\prime$) and ensemble average. The term $\bar{\pi
}^{xy}(t)$ represents the volume-averaged $xy$-component of the
energy-momentum tensor at time $t$: $\bar{\pi }^{xy}(t)=\sum
(p_{i}^{x}p_{i}^{y}/p^{0}_{i})/V$~\cite{Zhao:2020yvf}. 
The correlation function in the above relation is known to damp
exponentially with time~\cite{Muronga:2003tb,Demir:2008tr}:
\begin{equation}
\langle \bar{\pi }^{xy}(t+t^\prime)\,\bar{\pi }^{xy}(t^\prime) \rangle
=\langle \bar{\pi }^{xy}(t^\prime)\,\bar{\pi }^{xy}(t^\prime)
\rangle~e^{-t/\tau}, 
\label{exptau}
\end{equation}
where $\tau$ is the corresponding relaxation time. In addition, the 
average variance of $\bar{\pi }^{xy}$ in equilibrium is given by
$\langle \bar{\pi }^{xy}(t^\prime)\,\bar{\pi }^{xy}(t^\prime) \rangle
=4\epsilon \,T/(15V)$, where $\epsilon$ is the energy
density of the partons in equilibrium. One then has
\begin{equation}
\eta =\frac{4}{15}\, \epsilon \,\tau.
\end{equation}
In practice, we first extract the relaxation time $\tau$ from the
numerical calculation of the correlation function in
Eq.~(\ref{exptau}) using the ZPC parton
cascade~\cite{Zhang:1997ej,Zhao:2020yvf} and then obtain the shear
viscosity from the above relation. 

\section{Results for gluons in a box}

\begin{figure}[h!]
\includegraphics[scale=1]{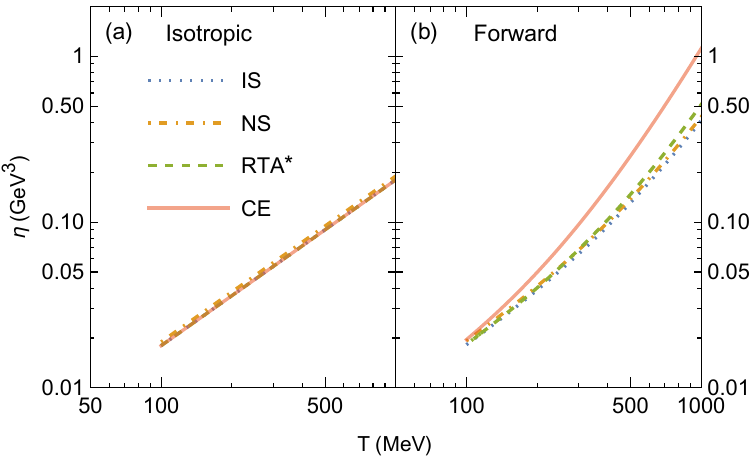}
\caption{The shear viscosity versus temperature from four analytical
  methods   for massless partons under (a) isotropic scatterings or
  (b) forward-angle scatterings, both at $\sigma=2.6$ mb with
  $\alpha_s=0.47$.  The methods include Israel-Stewart (IS),
  Navier-Stokes (NS),  modified relaxation time approximation
  (RTA$^*$), and  Chapman-Enskog (CE).}
\label{fig1}
\end{figure}

In this section, we consider a gluon gas in a box at a
given temperature with an elastic scattering cross section of
$\sigma=2.6$ mb and $\alpha_s=\sqrt{2}/3$~\cite{Zhao:2020yvf}, which
corresponds to $\mu \simeq 0.69$ GeV. Figure~\ref{fig1} shows the
shear viscosity versus temperature from the four analytical methods
under isotropic scatterings in panel (a) and the AMPT forward
scatterings in panel (b).  We see in panel (a) that the viscosity for
isotropic scatterings at a fixed cross section is linear in $T$ for
each method, and the results from the Israel-Stewart, modified
relaxation time approximation, and the Chapman-Enskog methods are all
the same. The results from the Navier-Stokes method is also very
close, only about 5\% higher.

For forward scatterings, however, we see in panel (b) that the
viscosity results from the four methods are not the same. At high
temperatures, the viscosity from the Chapman-Enskog method is much
higher than results from the other three methods that are relatively
close to each other. Also, the shear viscosity for forward
scatterings is significantly higher than that for isotropic
scatterings for every method; this is a result of the smaller
transport cross section as $h(a)<2/3$ for forward scatterings. At
low temperatures, the four methods give almost the same result because
$a \gg 1$ there, which makes the forward scattering cross section of
Eq.\eqref{dsdt} almost isotropic. Note that the results shown in
Fig.~\ref{fig1} apply not only to massless gluons but also to massless
partons.

\begin{figure}[h!]
\includegraphics[scale=1]{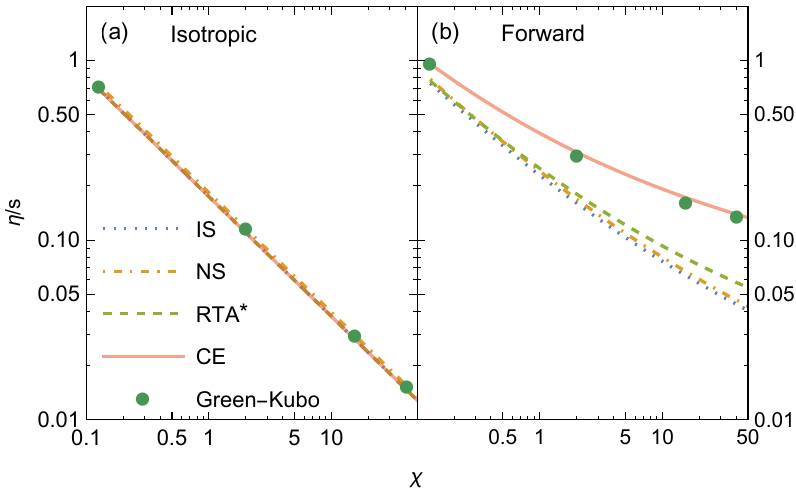}
\caption{The ratio of shear viscosity to entropy density versus the
  opacity parameter $\chi$ for massless gluons in equilibrium
 under (a) isotropic scatterings or (b) forward-angle scatterings of
 $\sigma=2.6$ mb with    $\alpha_s=0.47$. Results from four analytical
 methods (curves) are   compared with the numerical values from the
 Green-Kubo relation (circles).}
\label{fig2} 
\end{figure}

Results of the $\eta/s$ ratio for the gluon gas are shown versus the
opacity parameter  for isotropic scatterings in Fig.\ref{fig2}(a) and
the AMPT forward scatterings in Fig.\ref{fig2}(b).  The unitless
opacity parameter is defined as~\cite{Zhang:1998tj, Zhao:2020yvf} 
\begin{equation}
\chi =\sqrt{\frac{\sigma}{\pi } }/\lambda =n \sqrt{\frac{\sigma^3}{\pi } },
\label{chi}
\end{equation}
where $\lambda$ is the mean free path, and $n$ is the parton number
density. For a gluon gas under isotropic scatterings, we then have 
\begin{equation}
\left(\frac{\eta}{s}\right)^{\mathrm{IS}} \simeq
\frac{0.1744}{\chi^{2/3}},~~
\left(\frac{\eta}{s}\right)^{\mathrm{NS}}\simeq\frac{0.1839}{\chi^{2/3}}.
\end{equation}
We see in Fig.\ref{fig2}(a) that the $\eta/s$ ratios from the four
analytical methods are almost the same. They also agree well with the
circles, which represent our previous numerical
results~\cite{Zhao:2020yvf} obtained from the  Green-Kubo relation 
for four cases ($T=0.2$ GeV and $\sigma=2.6$ mb, $T=0.5$ GeV and
$\sigma=2.6$ mb, $T=0.7$ GeV and $\sigma=5.2$ mb,  $T=0.7$ GeV and
$\sigma=10$ mb).  Note that the Green-Kubo results shown in this
section represent the results using parton subdivision with a
subdivision factor $l=10^6$~\cite{Zhao:2020yvf} that essentially removes the
causality violation in parton cascade calculations.  

Figure~\ref{fig2}(b) shows the $\eta/s$ results for forward scatterings. 
Results from the Chapman-Enskog method agree well with the previous
numerical results from the Green-Kubo relation~\cite{Zhao:2020yvf},
while results from the other three analytical methods are too low. 
Note that the $\eta/s$ ratio for anisotropic scatterings is no longer
only a function of the opacity parameter $\chi$; it also depends on
the $\alpha_s$ value because the $h(a)$ function leads to a dependence
of $\eta$ on $\mu/T$, which is $\propto \alpha_s/\chi^{1/3}$.
At a fixed $\alpha_s$ value, which is the case for the calculations
shown in Fig.\ref{fig2}, the $\eta/s$ ratio is still a function of
$\chi$ only. Also note that it has been shown
earlier~\cite{Plumari:2012ep} that the Chapman-Enskog method agrees
well with the Green-Kubo results for forward scatterings. Therefore,
we shall mostly use the Chapman-Enskog viscosity in the following.

\section{Time Evolution of $\eta$ and $\eta/s$ of the parton matter from the AMPT model} 
\label{ampt}

We now consider the parton matter in the string melting version of the
AMPT model~\cite{Lin:2004en}.  
In an earlier study~\cite{Lin:2014tya}, we have calculated the time
evolution of the parton energy density $\epsilon$, number density, 
mean transverse momentum $\langle p_T\rangle$, and mean energy in the
center cell of central and midcentral Au+Au collisions at $200A$ GeV
and Pb+Pb collisions at $2760A$ GeV. 
Note that the study used the parton scattering cross section given by
Eq.\eqref{dsdt} with $\sigma=3$ mb and $\alpha_s=0.33$. 
We then extracted from each of these quantities the effective
temperature including $T_\epsilon$ and $T_{\langle p_T\rangle}$, 
which are found to be quite different; this also means that the parton
matter is in partial (not full) equilibrium~\cite{Lin:2014tya}. 
Partial chemical equilibrium can be characterized by the fugacity parameter
$\exp(\phi/T)$, where $\phi$ is the chemical potential. 
The shear viscosity for distributions in thermal equilibrium but with
a non-zero chemical potential has been calculated in the relaxation
time approximation~\cite{Chakraborty:2010fr,Wiranata:2012br}, 
and the viscosity can be shown to be independent of the fugacity 
for Boltzmann distributions. 
Therefore, the $\eta$ expressions shown in the previous
Section can be applied to a parton matter under partial chemical
equilibrium.

\begin{figure}[h!]
\includegraphics[scale=0.7]{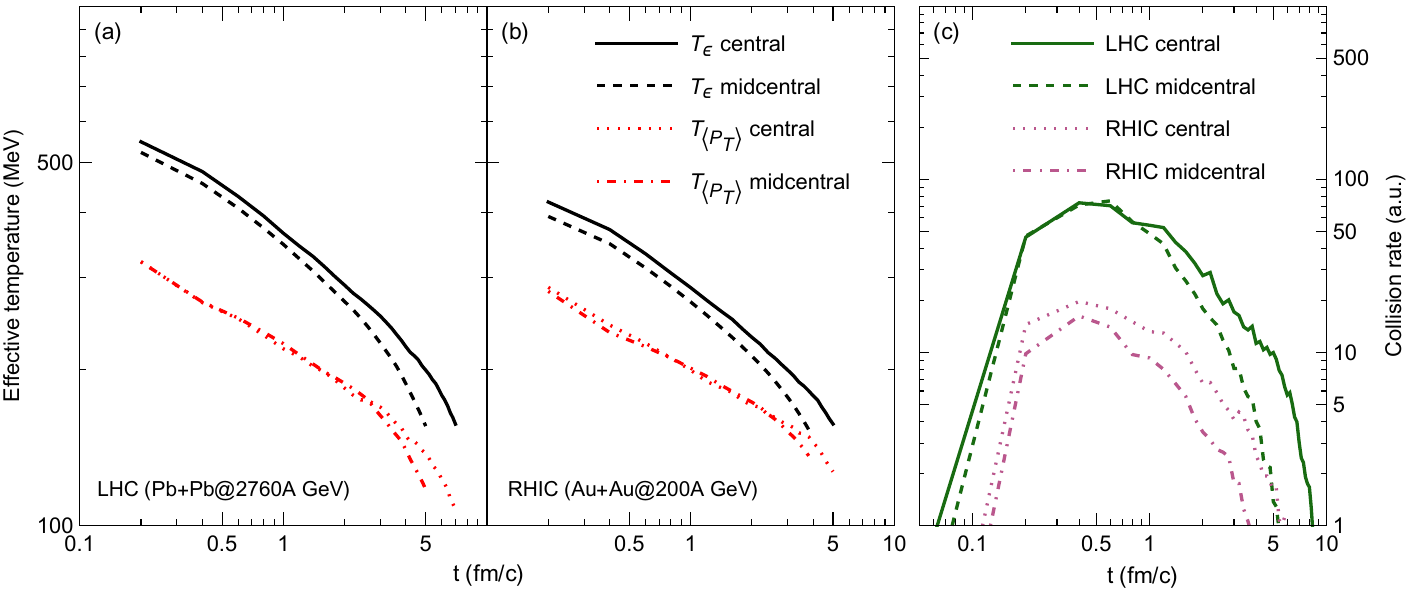}
\caption{The time evolutions of the effective temperatures $T_\epsilon$
  and  $T_{\langle p_T\rangle}$ for the center cell in (a) Pb+Pb
  collisions at $2760A$ GeV and (b) Au+Au collisions at $200A$
  GeV. (c) shows the time evolutions of the collision rates (in
  arbitrary units) for the center cell.}
\label{fig3} 
\end{figure}

Figure~\ref{fig3} shows the time evolutions of two effective
temperatures obtained earlier~\cite{Lin:2014tya} from the AMPT model for
the center cell of central and midcentral Pb+Pb collisions at $2760A$
GeV in panel (a) and Au+Au collisions at $200A$ GeV in panel (b). 
Central and midcentral collisions at the RHIC energy refer to Au+Au
events with $b<3$ fm and $b=7.3$ fm, respectively; while those at the
LHC energy refer to Pb+Pb events with $b<3.5$ fm and $b=7.8$ fm,
respectively. The center cell is the volume within
mid-spacetime-rapidity in the center of the transverse plane within
$|x|<1/2$ fm and $|y|<1/2$ fm. 
Note that we have plotted the time evolutions up to the time when the 
center cell reaches $T_\epsilon=155$ MeV, approximately the
temperature of the QCD crossover transition at zero net-baryon
density. The effective temperature $T_\epsilon$ is determined from the
parton energy density~\cite{Lin:2014tya} assuming that the parton
matter is a QGP with massless gluons and (anti)quarks of three flavors
under the Boltzmann statistics, i.e., using $\epsilon=3g_{\rm 
  B}T_\epsilon^4/\pi^2$ with $g_{\rm B}=52$.  On the other hand,
temperature $T_{\langle   p_T\rangle}$ is determined from the mean
parton transverse momentum for massless partons under the Boltzmann
statistics, i.e., using $\langle p_T\rangle=3\pi T_{\langle
  p_T\rangle}/4$. In full thermal and chemical equilibrium, these two
temperature values are the same. However, we see from
Fig.~\ref{fig3}(a) and (b) that they are very different for the parton
matter from the AMPT model, with $T_\epsilon>T_{\langle p_T\rangle}$
for the four collision systems. 
Note that in principle the parton matter is not in full thermal
equilibrium because the pressures along different axes are not 
isotropic during the evolution of the parton matter in the AMPT 
model~\cite{Lin:2014tya,Wang:2021owa}.  This causes the effective
temperatures extracted from ${\langle   p_T\rangle}$, ${\langle
  p\rangle}$, and ${\langle p_T^2\rangle}$ to be
different~\cite{Lin:2014tya}. On the other hand, the difference
between $T_\epsilon$ and  $T_{\langle   p_T\rangle}$ is typically much
larger~\cite{Lin:2014tya}. Therefore, in this section we approximate
the parton matter in the center cell as a QGP in full thermal
equilibrium (at temperature $T_{\langle p_T\rangle}$) but partial
chemical equilibrium (with the energy density given by temperature
$T_\epsilon$) so that we can apply the analytical expressions 
for the shear viscosity. Note that this corresponds to a QGP with the
fugacity of $(T_\epsilon/T_{\langle p_T\rangle})^4$. 

\begin{figure}[h!]
\includegraphics[scale=0.85]{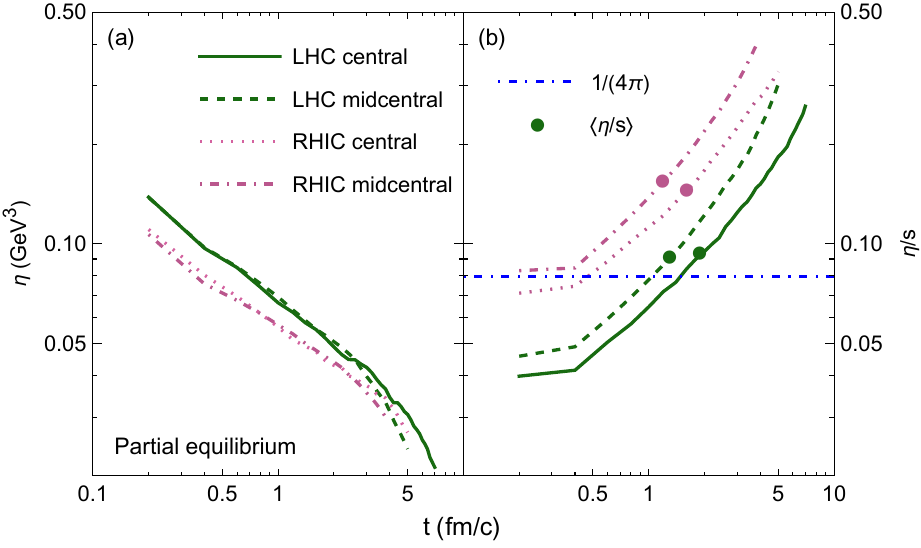}
\caption{The time evolutions of (a) the CE shear viscosity and (b)
  $\eta/s$ of the center cell in Pb+Pb   collisions at $2760A$ GeV
  and Au+Au collisions at  $200A$  GeV. Here the parton matter from
  the AMPT model is considered as a QGP in partial equilibrium, where
  the momentum distribution corresponds   to temperature   $T_{\langle
    p_T\rangle}$ but the energy density corresponds to   temperature
  $T_\epsilon$. Circle on each curve represents the   corresponding
  average $\eta/s$ weighed by the collision rate.}
\label{fig:partial}
\end{figure}

Figure~\ref{fig:partial}(a) shows the time evolutions of the shear
viscosity from the CE method for the parton matter in the center cell
in partial equilibrium. Since the shear viscosity is determined by the 
momentum transfer but not fugacity, it should be calculated with
temperature $T_{\langle   p_T\rangle}$ that represents the momentum
distribution.  As expected for a constant cross section, the shear
viscosity decreases with time as the parton matter cools down.  The
time evolutions of $\eta/s$ are shown in
Fig.~\ref{fig:partial}(b). Note that throughout this Section the
entropy density for the parton matter from the AMPT model is taken as
the full  equilibrium value, $s_{eq}=4 g_{\rm
  B}T_{\epsilon}^3/\pi^2$, i.e., the value when the parton matter is  
assumed to be a QGP in full equilibrium with energy density
$\epsilon$. We see in Fig.~\ref{fig:partial}(b) that the $\eta/s$
ratio strongly increases with time as the effective temperature
decreases~\cite{Xu:2011fi}.  Also, the $\eta/s$ ratio is lower at the
LHC energy than the RHIC energy and lower in central events than
semicentral events.  These features are mainly because of the higher
entropy density at higher temperatures ($s \propto T^3$) and 
$\eta \propto T/g_2(w)$, where $1/g_2(w)$ grows with $T$ slower than
$\sim T^{1.4}$ within the relevant temperature range (as can be
observed from the CE curves in Fig.\ref{fig1}).  In addition, at very
early times the $\eta/s$ values from the AMPT model are sometimes
significantly below $1/(4\pi)$, the lower bound from the conformal
field theory~\cite{Kovtun:2004de}.  This is partly a result of the
high parton density (or equivalently the high $T_{\epsilon}$) in the
string melting AMPT model despite the small parton cross section.

The $\eta/s$ ratio in Fig.~\ref{fig:partial}(b) 
decreases as the temperature increases; 
this temperature dependence is opposite to that 
preferred by the Bayesian analysis of heavy ion experimental data
~\cite{JETSCAPE:2020mzn}, where the preferred $\eta/s$ 
increases as the temperature increases. 
This ``wrong'' temperature dependence is a result of using a constant
parton cross section. 
If we use in Eq.\eqref{dsdt} a temperature-dependent Debye mass as 
$\mu \propto g T$ with $g$ being the QCD gauge coupling, 
we would have $\sigma \propto \alpha_s/T^2$ and thus $\eta^{\mathrm{CE}}/s
\propto 1/(\alpha_s g_2)$ would increase with $T$. This would be 
qualitatively similar to earlier pQCD
studies~\cite{Arnold:2000dr,Arnold:2003zc} that used temperature-dependent
Debye screening masses for $2\leftrightarrow 2$ and $1\leftrightarrow
2$ parton processes,  which showed that the $\eta/s$ of the QGP
increases with the temperature~\cite{Csernai:2006zz}. 

Our early study on the effective temperatures of the parton
matter~\cite{Lin:2014tya} found that $\sigma=3$ mb can roughly reproduce 
the bulk observables in the four collision systems, including the pion
and kaon elliptic flows at low $p_T$.  
An effective $\eta/s$ value averaged over its time evolution is a good
overall measure of the property of the parton matter. 
Since parton scatterings convert the initial spatial geometry into
anisotropic flows including the elliptic flow~\cite{Lin:2001zk,He:2015hfa}, 
we use the collision rate in the center cell as the weight in the
average. From Fig.~\ref{fig3}(c), we see that the collision rate
(i.e., the number of parton collisions per time in the center cell)
rises at early times before it decreases; the rise is because a parton
is only allowed to interact a finite formation time
after it is produced from the collision of the two nuclei. 
The circle on each curve in Fig.~\ref{fig:partial}(b) gives the
average $\eta/s$ value, which is very small and ranges from just above
$1/(4\pi)$ to about twice $1/(4\pi)$.

\begin{figure}[h!]
\includegraphics[scale=0.8]{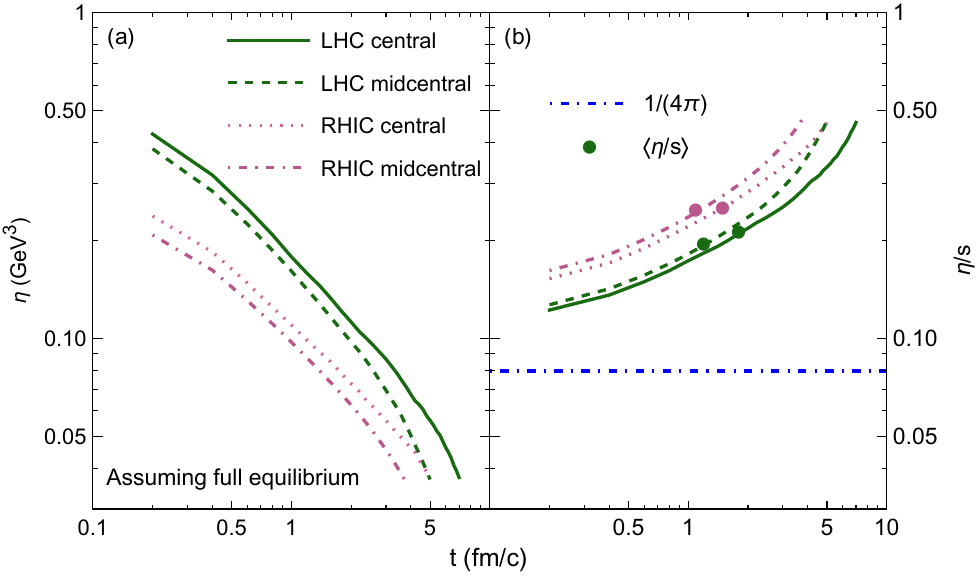}
\caption{Same as Fig.\ref{fig:partial}, but the parton matter is assumed to be a
QGP in full equilibrium at temperature $T_\epsilon$.}
\label{fig:full}
\end{figure}

Because the AMPT model lacks inelastic parton processes such as $2
\leftrightarrow 3$ processes~\cite{Xu:2004mz}, its parton matter 
cannot approach chemical equilibrium. 
We thus also calculate the shear viscosity and $\eta/s$ by assuming
that the parton matter is in full (chemical and thermal) equilibrium
with the same energy density (i.e., at temperature $T_\epsilon$). 
Figure~\ref{fig:full} shows the full equilibrium $\eta$ in panel (a)
and the $\eta/s$ ratio in panel (b), 
where $\eta$ is calculated using temperature $T_\epsilon$ (instead of
$T_{\langle p_T\rangle}$ for the partial equilibrium case). 
Compared to Fig.~\ref{fig:partial}(a), we see that the full
equilibrium viscosity is significantly higher. This is because
$T_\epsilon>T_{\langle p_T\rangle}$, and note that the increase of
$\eta$ with temperature is faster than linear for anisotropic
scatterings due to the $1/g_2(w)$ term in Eq.\eqref{ce2}. 
Compared to the partial equilibrium case, 
the $\eta/s$ ratios for the full equilibrium case are significantly
higher, and they are above $1/(4\pi)$ in all times for the four
collision systems. As shown by the circles in Fig.~\ref{fig:full}, the
averaged $\eta/s$ values are also significantly higher than the
partial equilibrium case, around three times $1/(4\pi)$ for the four
systems.  Note that the partial equilibrium nature of the AMPT parton
matter, particularly the low $T_{\langle p_T\rangle}$ relative to the
high energy density, helps the model to reproduce the large observed
elliptic flow~\cite{Lin:2001zk,Molnar:2019yam}; we now know that this
is related to its lower shear viscosity and $\eta/s$ than the full
equilibrium case.

\section{Discussions}

\begin{figure}[h!]
\includegraphics[scale=0.8]{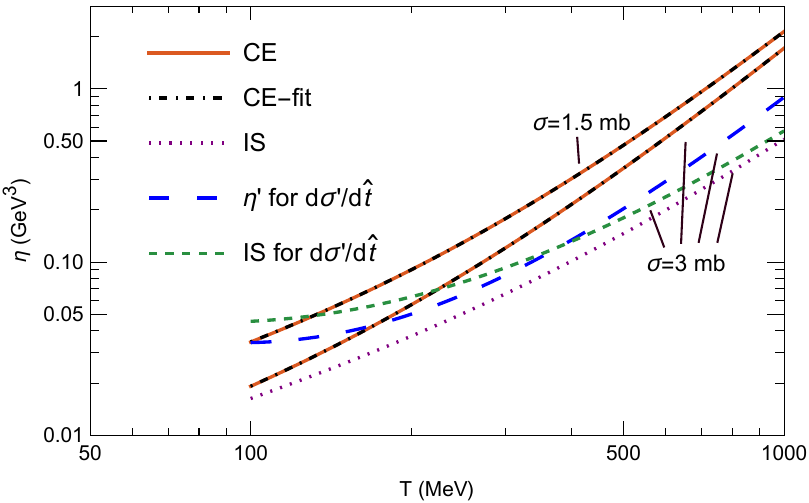}
\caption{The shear viscosity versus temperature for a QGP in full equilibrium
under forward-angle scatterings at $\sigma=3$ mb with 
$\alpha_s=0.33$. Results from the IS method, the CE method, and a fit
of the CE result are shown, together with the estimated shear
viscosity from Ref.~\cite{Xu:2011fi} (long dashed curve) and 
the IS result (dashed curve) for $d\sigma^\prime/d{\hat t}$. 
The CE results for $\sigma=1.5$ mb are also shown.}
\label{fig:compare}
\end{figure}

Since the CE viscosity of Eq.\eqref{ce2} contains the $g_2(w)$
function that involves an integral, we have fit it with the following 
to make the calculation of $\eta^{\mathrm{CE}}$ easier for the parton
cross section of Eq.\eqref{dsdt}:
\begin{equation}
g_2^{\mathrm{fit}}(w)=h\!\left(\frac{w^2}{v^2} \right), 
{\rm~~with~~}v=11.31-4.847\exp (-0.1378~w^{0.7338}).
\end{equation}
Figure~\ref{fig:compare} shows that the viscosity values using the
fit function (dot-dashed curves) overlap with the exact CE results
(solid curves) regardless of the cross section value.

The AMPT model results in Sec.\ref{ampt}  have been obtained at 
$\sigma=3$ mb, because this value enables the string melting version of
the AMPT model to approximately reproduce the elliptic flow
data at low $p_T$~\cite{Lin:2001zk,Lin:2014tya}. 
In the AMPT model with the new quark coalescence that allows a parton
to have the freedom to form either a meson or a
baryon~\cite{He:2017tla}, a smaller parton cross section of
$\sigma=1.5$ mb is found to approximately reproduce the elliptic flow 
data. In that case, the shear viscosity would be higher (by a factor
between one and two), as shown by the two solid curves in
Fig.~\ref{fig:compare}. 

An earlier study~\cite{Xu:2011fi} used the IS method to estimate 
the shear viscosity and  $\eta/s$ of the parton matter under forward
scatterings of the AMPT model. A recent study~\cite{Magdy:2021sip}
used the same estimates. They wrote the differential cross
section as
\begin{equation}
\frac{d\sigma^\prime}{d{\hat t}}=\frac {9\pi\alpha_s^2}{2}
\frac{1}{({\hat t}-\mu^2)^2}, 
\end{equation}
i.e., the same as Eq.\eqref{dsdt} except for the factor 
$(1+a)$. Note that the total cross section in this case would
be given by  $\sigma^\prime=\sigma/(1+a)$~\cite{Lin:2004en}, which 
depends on energy or temperature. They further approximated the
thermal average of the transport cross section as 
$\langle \sigma^\prime h(a) \rangle
\simeq \sigma^\prime h(a) |_{a \rightarrow \mu^2/\langle {\hat s} \rangle}$, 
where $\langle {\hat s} \rangle=18T^2$. 
The shear viscosity is then estimated as~\cite{Xu:2011fi}
\begin{equation}
\eta^\prime \simeq \frac{4~T^3}
{5\pi\alpha_s^2\left [ \left(1+\frac{\mu^2}{9T^2}
    \right)\ln\left(1+\frac{18T^2}{\mu^2} \right)-2 \right ]}.
\end{equation}
This estimated viscosity $\eta^\prime$ for $\sigma=3$ mb and
$\alpha_s=0.33$ is shown as the long dashed curve in
Fig.\ref{fig:compare}. When we perform the proper thermal average of
the transport cross section instead of approximating it with $a
\rightarrow \mu^2/\langle {\hat s} \rangle$, we get the IS viscosity
for $d\sigma^\prime/d{\hat t}$ as shown by the dashed curve in
Fig.\ref{fig:compare}. 
We see that for $d\sigma^\prime/d{\hat t}$ the estimated $\eta^\prime$
is higher (or lower) than the IS curve at high (or low) temperatures, 
reflecting the inaccuracy of the approximation of the thermal average in
$\eta^\prime$. 
We also see that the IS curve for $d\sigma^\prime/d{\hat t}$ approach
the IS curve for $d\sigma/d{\hat t}$ of Eq.(\ref{dsdt}) (dotted curve)
at high temperatures since  $(1+\mu^2/{\hat s}) \sim 1$ there, 
while at low temperatures $\eta^{\mathrm{IS}}$ for
$d\sigma^\prime/d{\hat t}$ is much higher due to its smaller total
cross section. 

In this study we have only used the first-order Chapman-Enskog
expression for the shear viscosity, which is
$\eta^{\mathrm{CE}}=(6/5)T/\sigma$ for isotropic
scatterings. Higher-order CE terms have been 
derived earlier~\cite{Wiranata:2012br} and shown to only have a small
effect. For example, for massless particles under isotropic
scatterings, including the second-order term changes the $\eta^{\mathrm{CE}}$
front coefficient from $6/5$ to 1.256 while including terms up to the 16th
order only changes the front coefficient to 1.268. 
Note that there is a typo in Eq.(25) of Ref.\cite{Wiranata:2012br} 
for the second-order CE term, where $\gamma_0^2c_{00}$ should be
$\gamma_0^2c_{11}$; this typo has been corrected in
Ref.~\cite{Plumari:2012ep}. 

\section{Conclusion}

In this study, we investigate the shear viscosity 
and the $\eta/s$ ratio of a massless parton matter under isotropic or
forward-angle two-body scatterings. 
We first compare the analytical results from the Israel-Stewart,
Navier-Stokes, relaxation time approximation, and Chapman-Enskog
methods with the numerical results from the Green-Kubo relation.  
We confirm the earlier finding that only the Chapman-Enskog
method agrees well with the numerical results for anisotropic
scatterings. 
We then apply the Chapman-Enskog method to calculate the time
evolution of the shear viscosity and $\eta/s$ of the center cell of
the parton matter from the string melting  AMPT model for central and
midcentral Au+Au collisions at $200A$ GeV and Pb+Pb collisions at
$2760A$ GeV.  
We approximate the parton matter as a quark-gluon
plasma in thermal equilibrium at effective temperature
$T_{\langle p_T\rangle}$ but in partial chemical equilibrium, where
the energy density  corresponds to a higher effective temperature
$T_\epsilon$. Due to the partial equilibrium nature, we use
temperature $T_{\langle p_T\rangle}$ to calculate the shear viscosity,
which has a lower value than the case if the parton matter were in
full equilibrium at temperature $T_\epsilon$. 
For the AMPT model that approximately reproduces the elliptic flow
data at low transverse momentum with a parton cross section $\sigma=3$
mb, the average $\eta/s$ for each of the four collision systems 
is found to be very small, between one to two times $1/(4\pi)$. 
If the parton matter were in full thermal and chemical equilibrium,
the average $\eta/s$ would be higher, around three times
$1/(4\pi)$. In addition, the $\eta/s$ ratio from the current model
decreases as the temperature increases, contrary to pQCD results that
use temperature-dependent Debye masses. This is a result of the AMPT
model using a constant Debye mass for parton  scatterings, 
which could be an area for future improvements. 

\section*{Acknowledgements}

This work is supported by the National Science Foundation under Award
No. PHY-2012947.

\end{document}